\documentclass[aip,pop,superscriptaddress,preprint]{revtex4-2}
\usepackage{amsmath}
\usepackage{graphicx}
\usepackage{color}
\usepackage{natbib}
\usepackage{subcaption}
\usepackage{bmpsize}
\begin{document}

\title{On How Avalanches Penetrate the SOL and Broaden Heat Loads}

\author{Y. Kosuga}
\affiliation{Research Institute for Applied Mechanics, Kyushu University, Fukuoka, 816-8580, Japan}

\author{R. Matsui}
\affiliation{Interdisciplinary Graduate School for Engineering Sciences, Kyushu University, Fukuoka, 816-8580, Japan}

\author{P.H. Diamond}
\affiliation{University of California, San Diego, La Jolla, CA, 92093, USA}

\begin{abstract}

Recent experiments reported a correlation between power law core temperature spectra and $D_\alpha$ emission, suggesting that heat avalanches penetrate the SOL. This paper  derives a threshold criterion for avalanche penetration using a reduced model. Avalanches with $(\nabla\tilde T)_{rms}>\nabla\tilde T_{crit}$ at the separatrix are predicted to penetrate, and so broaden the SOL and heat load distribution. $\nabla\tilde T_{crit}$ is $\sim 1/\tau_\parallel$, where $\tau_\parallel$ is the parallel heat flow time through the SOL. Penetration occurs when avalanches are strong enough to steepen sufficiently to shock at the separatrix. A positive correlation is found between the nonlinear drive for steepening and the penetration depth. In particular, penetration depth exceeds that of the heuristic drift limit when shocks form. Implications for numerical and physical experiments are also discussed.

%Role of avalanches in broadening SOL width is addressed. A simplified model is introduced to describe avalanche penetration into SOL. The model is an extension of Burgers model for avalanche propagation in main plasmas, to include additional loss along the open field line and magnetic drifts. A condition is derived for avalanche penetration  via a shock formation criterion. The extended model is numerically solved to demonstrate the penetration of avalanches into SOL. A positive correlation is found between the nonlinear drive for shock and the penetration depth. Implications on numerical and physical experiments are also discussed.

\end{abstract}

\maketitle

\section{Introduction}
The problem of divertor heat load is one of the most critical issues that magnetic fusion research confronts today\cite{ITERRevCh4}. 
%The main plasma is heated to produce fusion reaction, which ultimately leads to burning plasmas. This heat stored in the system, however, needs to be eventually exhausted. They are released from the confined plasma mainly through transport, and enter the scrape off layer (SOL) region. Then it propagates along the magnetic field, which reaches divertor plates in the end. 
From engineering requirements\cite{SOLHeatLoad}, the heat flux to the divertor plate in future experiments, such as ITER, must not exceed $\sim 20\mathrm{[MW/m^2]}$, which corresponds to a SOL width in the range of $\lambda_T\sim 5\mathrm{[mm]}$. However, there are studies that raise concerns, pointing out that the SOL width in ITER may be narrower than required\cite{EichScaling,Goldston}. Those studies estimate the SOL width by extrapolating an experimental scaling\cite{EichScaling}, which is explained by the heuristic drift model\cite{Goldston}. Note that the latter work does not address the role of fluctuations in the SOL, since it is assumed that the strong SOL $E\times B$ shear in H-mode will suppress local instabilities there.

While the role of turbulence is dismissed in the above, there are several studies\cite{XuBoutHeat,SOLLocalTurb,YK15,TurbSOLHL-2A,SpreadingDIII-D} that report the positive correlation between the SOL width and turbulent fluctuations at the edge-pedestal\cite{TurbSOLHL-2A,SpreadingDIII-D} and at SOL\cite{TurbSOLHL-2A}. Indeed, even if local modes in the SOL are suppressed, SOL fluctuations can be energized by pedestal turbulence. This is observed in experiments\cite{SpreadingLHD,SpreadingDIII-D,SpreadingHL-2A}. In some cases, the flux of fluctuation energy from pedestal to SOL is suffficient to drive fluctuation activity in the edge region\cite{ManzSpreading,SpreadingDIII-D}. The flux of fluctuation energy through the LCFS emerges as a critical parameter for SOL widening, as indicated by a simplified model\cite{ChuSpreadingWidth} and numerical simulations based on BOUT++\cite{SpreadingELMSOL,SpreadingWidthBout_DIII-D}.

As indicated above, turbulence in the boundary region is unique in that it can be supported, not only by locally unstable modes, but by turbulence spreading phenomena\cite{Boedo1,BlobSergei,GKSOLWidth}. A manifestation of these is an avalanche\cite{JKPSMesoReview}, which is a propagating sequence of temperature corrugations, leading to bursty transport events. Avalanches are observed both in numerical\cite{SOCRealization,GarbetAva1,GYSELAAvalanches,GT5D} and physical experiments. In physical experiments, avalanches are often seen as electron temperature corrugations via electron cyclotron emission (ECE), as reported from DIII-D\cite{PolitzerAvalanches}, KSTAR\cite{KSTARMesoscopic}, JT-60U\cite{AvalancheHeliotronJT60}, Heliotron-J\cite{HeliotronAvalancheSOL}, etc. In particular, recent observation on Heliotron-J\cite{HeliotronAvalancheSOL} indicates the penetration of avalanches into SOL plasmas. In this experiment, avalanches originate in the core region, as detected in ECE data. Avalanche propagation is also quantified by calculating the radial correlation of the event. The correlation remains finite outside of the LCFS, indicating that avalanches penetrate the SOL. The footprint of fluctuation activity also appears on the $D_\alpha$ signal and in recent probe measurements.  Interestingly, penetration is only observed for a high Power experiment, suggesting there may be a critical condition for the avalanche penetration. However, the mechanism for avalanche penetration remains unclear. Moreover, extent of SOL broadening also needs to be addressed.

% The core avalanches triggers fluctuation activity in SOL region, as indicated by the $D_\alpha$ signal and probe measurements. 

% where the radial correlation of the event remains finite in the SOL region. The footprint of fluctuation activity also appears on the $D_\alpha$ signal and in recent probe measurements. Interestingly, the experiment on Heliotron-J reports that the penetration is only observed for a high Power experiment, suggesting there may be a condition for the avalanche penetration. However, a mechanism for avalanche penetration remains unclear. Moreover, implications for SOL broadening also need to be addressed.

%In this experiment, avalanche penetration is observed only for high input power. In order to understand behavior, physics needs to be clarified. Implications on the SOL broadening needs to be addressed, as well.

%avalanches are found to penetrate into the SOL region for high heating power. 

%The results from Heliotron-J indeed raise several questions. In particular, the experiment finds that there seems to be a condition for the avalanche penetration. In this experiment, avalanches are found to penetrate into the SOL region for high heating power. 

\begin{figure}[thb]
\centering\includegraphics[width=0.45\textwidth]{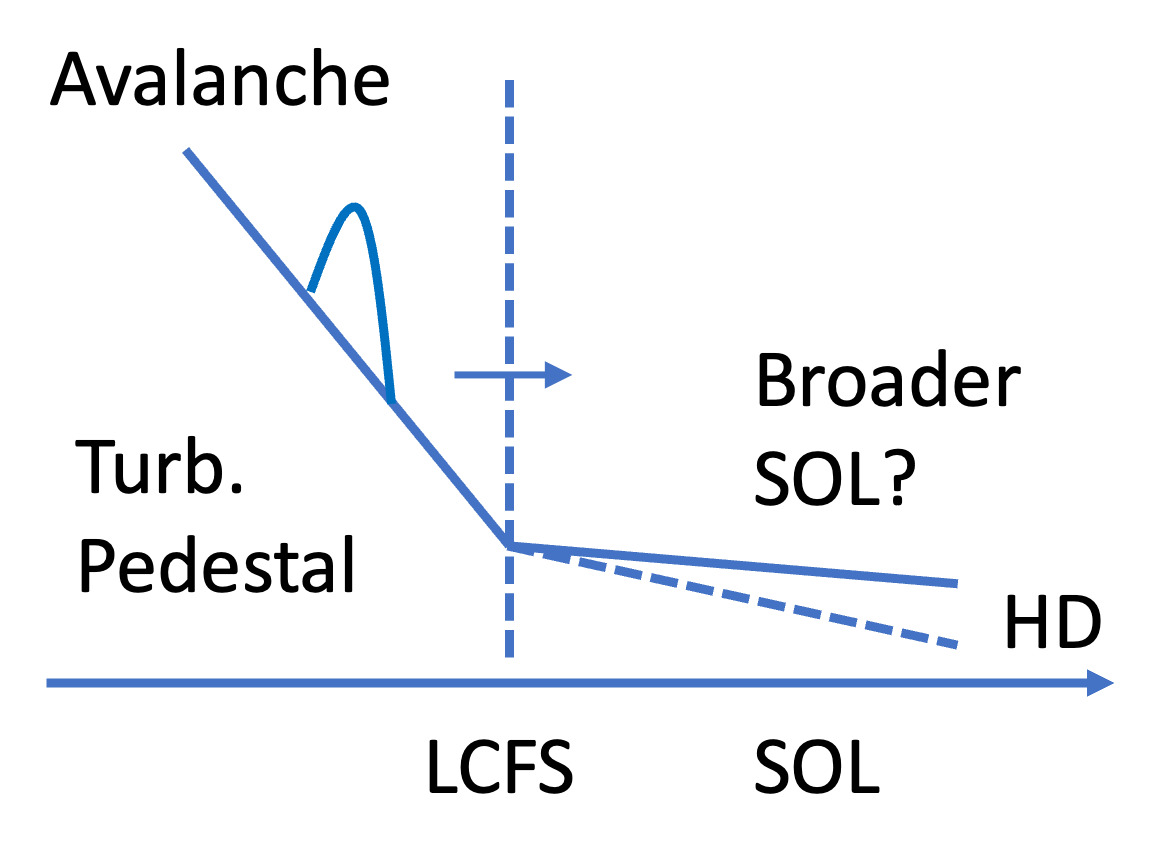}
\caption{A schematic picture of non-local drive of turbulence in the SOL. Avalanches and turbulence can propagate from the main plasma to reach the LCFS and to penetrate and broaden the SOL.}
\label{Fig:Penetration}
\end{figure}

\begin{figure}[thb]
\centering\includegraphics[width=0.45\textwidth]{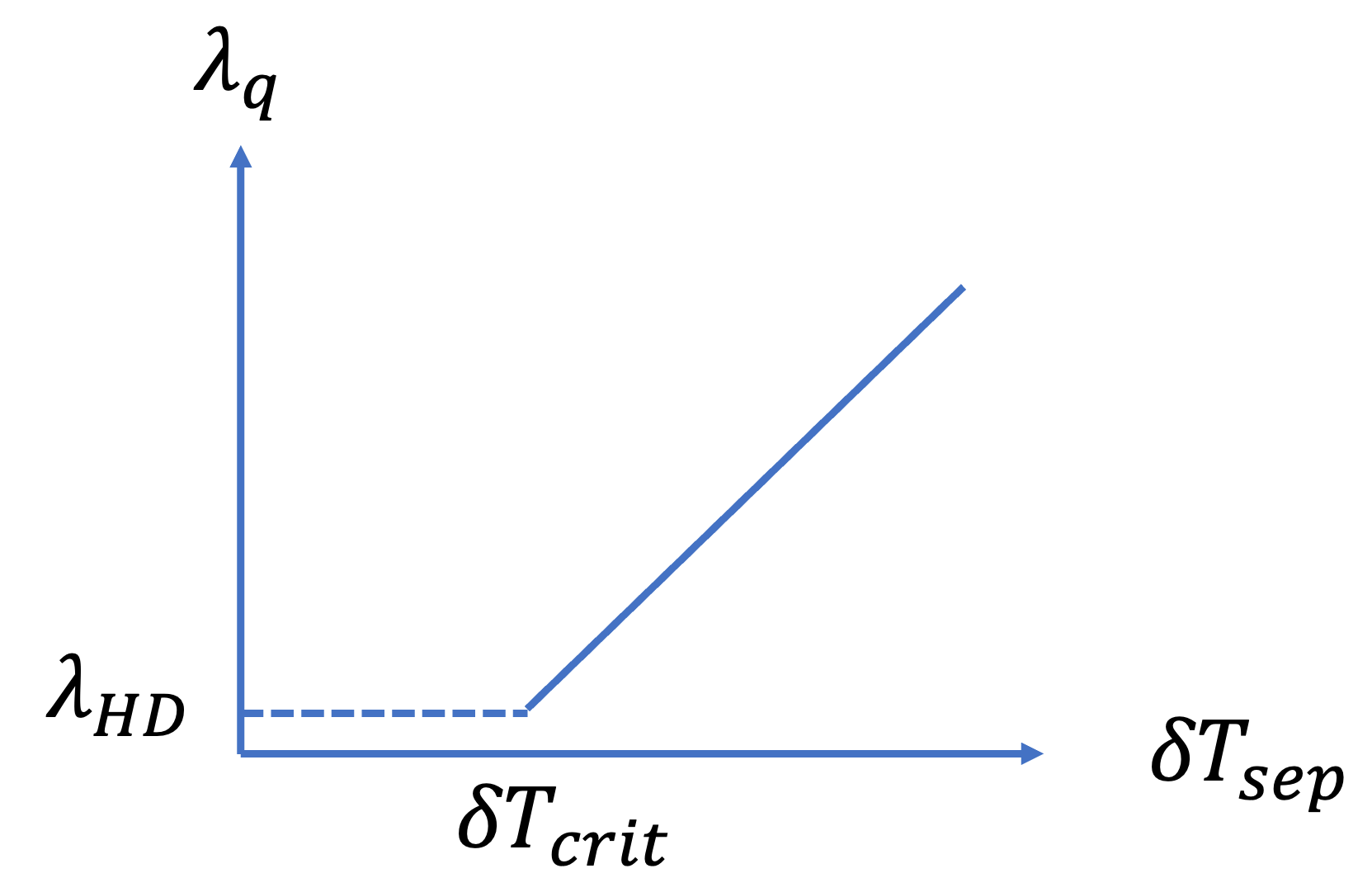}
\caption{A schematic picture of the dependence of the heat flux SOL width on the root-mean square value of the temperature perturbation at the LCFS. When the temperature perturbation at the LCFS exceeds a critical value, the width can exceed that of the heuristic drift model.}
\label{Fig:WidthTemp}
\end{figure}

The purpose of this work is then to elucidate the physics of penetration of avalanches into the SOL region and to quantify their role in broadening the SOL (Fig.\ref{Fig:Penetration}). We present a model to describe avalanche propagation in the SOL region\cite{AvalancheSOLModel}. The model extends the hydrodynamic model of avalanches using joint reflection symmetry, by including additional effects relevant in the SOL, such as the parallel damping. The extended model is formulated in terms of damped Burgers equation for temperature perturbation. Then the temperature perturbation is no longer a conserved order parameter. Thus, avalanches naturally attenuate and so require a critical drive for penetration. This condition for penetration is derived via a shock formation condition for the damped Burgers equation.  The equation is then solved numerically to demonstrate the penetration of avalanches into the SOL. We describe the dynamics of a single isolated pulse in the region and the relevance of boundary forcing, which mimics the dynamical avalanches bombarding the LCFS. The analysis identifies the root mean square part of the temperature gradient at the separatrix as the critical parameter for the penetration (Fig.\ref{Fig:WidthTemp}). The penetration depth is quantified, and the positive correlation is found between penetration depth and the nonlinear drive for shock formation. 

The remainder of the paper is organized as follows. Section 2 presents the model used in this work. Here we introduce a model for avalanches, and discuss an extension to include SOL-relevant effects. Section 3 is for numerical analysis. Section 4 is a summary and discussion.

\section{Model}

\begin{figure*}[htb]
\centering
\begin{subfigure}[b]{0.45\textwidth}
\includegraphics[width=\textwidth]{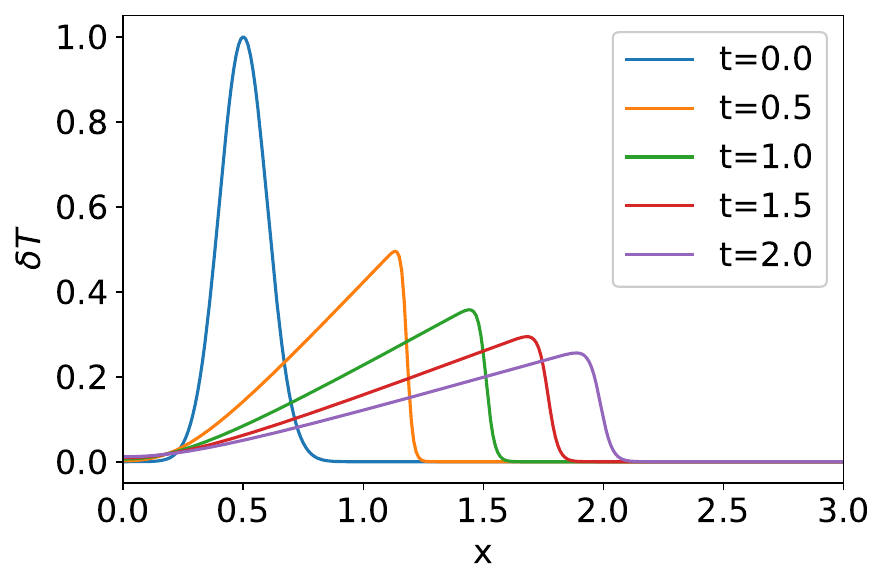}
\caption{Outward propagation of a blob}
\end{subfigure}
\centering
\begin{subfigure}[b]{0.45\textwidth}
\includegraphics[width=\textwidth]{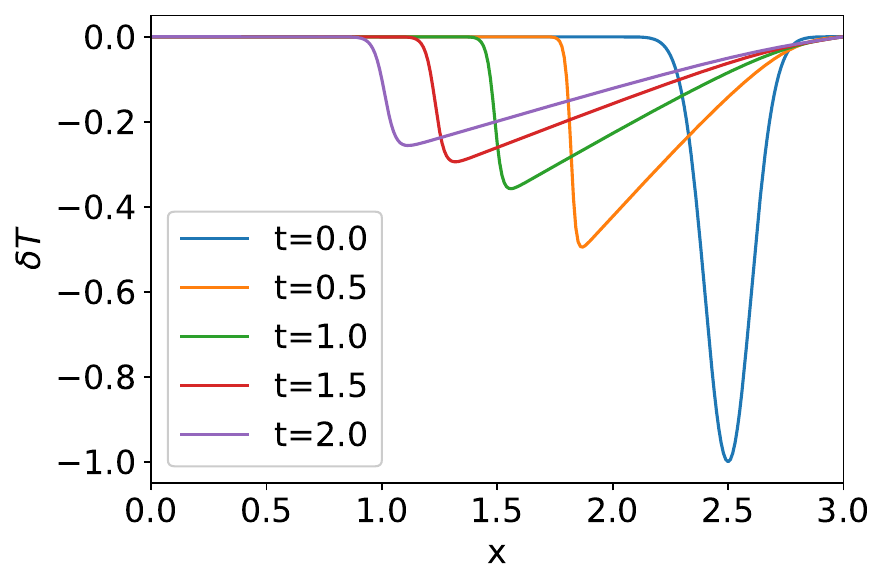}
\caption{Inward propagation of a void}
\end{subfigure}
\caption{Demonstration of Burgers equation having solutions with joint reflection symmetry}
\label{Fig:BlobVoid}
\end{figure*}

In this work, we are interested in the penetration of avalanches into the SOL. Here avalanches are sequential propagation of a meso scale temperature perturbation (or corrugation). The temperature profile dynamically fluctuates and deviates from a self-organized profile. This deviation amplifies the local temperature gradient and initiates the propagation of avalanches. The dynamics of avalanches can be formulated in the context of a hydrodynamic description. Here we are interested in the dynamics of the temperature deviation from a mean profile, $\delta T=T-T_0$. $T$ is the total temperature profile, and $T_0$ is a self-organized mean profile. The temperature deviation evolves according to the heat balance equation,
\begin{equation}
\partial_t\delta T+\partial_x Q[\delta T]=\chi_0\partial_x^2\delta T+\tilde{s}
\end{equation}
$\chi_0$ is a background conductivity (usually collisional or neoclassical conductivity) and $\tilde{s}$ is noise. $Q[\delta T]$ is the heat flux associated with avalanches. The form of $Q[\delta T]$ is constructed via a symmetry argument - more precisely, by exploiting the joint reflection symmetry(JRS)\cite{HK92,DHSOC}. JRS requires the total heat flux to be down the gradient. This allows two types of propagation, one is an out-going blob and the other is an in-coming void. In order to have both types of solutions, the equation needs to be invariant under the dual transformation of $x\to-x$ and $\delta T\to -\delta T$. The simplest form of the flux of JRS is given by
\begin{equation}
Q=\frac{\alpha}{2}\delta T^2-\chi_2\partial_x\delta T
\end{equation}
Then the evolution of temperature perturbation is given by Burgers model,
\begin{equation}
\partial_t\delta T+\alpha\delta T\partial_x\delta T=\chi_0\partial_x^2\delta T+\tilde{s}
\label{Eq:Burgers}
\end{equation}
$\chi_2$ is absorbed into $\chi_0$. Equation (\ref{Eq:Burgers}) describes avalanching turbulence as shock turbulence. Equation (\ref{Eq:Burgers}) indeed admits an out-going blob and an incoming void as its solution, as depicted in Fig.\ref{Fig:BlobVoid}. Here a localized pulse is initiated in the gaussian form, $A_0\exp(-(x-x_0)^2/(2\Delta^2))$, with $A_0=\pm1$, $x_0=0.5/2.5$, $\Delta =0.1$, $\chi_0$=0.01, $\alpha=1$ for a blob/void. Shock formation is also evident here.

Several relevant physics effects can be incorporated into avalanche dynamics. In this work, we are interested in avalanche penetration into the SOL plasmas, and discuss an extension of Burgers model for the SOL. One very important effect is the loss of heat due to flow along magnetic field. Unlike the main plasmas, where the magnetic field is closed, SOL plasmas support an additional loss channel due to heat flow along the magnetic field. This also applies to avalanches. While propagating in the radial direction, the temperature perturbation also propagates in the parallel direction and finally is lost to divertor plates. In this work, we employ a simplified description of the loss, as a Krook model for damping with the damping rate set by $1/\tau_\parallel$. In principle, the damping can be set either by $\sim c_s/l_\parallel$ in the sheath limited regime or by $1/\tau_\parallel\sim \chi_{e\parallel}/l_\parallel^2$ in the conduction limited regime. Here $c_s$ is the sound speed, $l_\parallel\sim qR$ is the typical length of the magnetic field, and $\chi_{e\parallel}$ is the electron heat conductivity in the parallel direction. We do not distinguish these here, and model the parallel loss with a single damping parameter $1/\tau_\parallel$ for simplicity. The evolution of $\delta T$ is then given by
\begin{equation}
\partial_t\delta T+v_{D}\partial_x\delta T+\alpha\delta T\partial_x\delta T=\chi_0\partial_x^2\delta T-\frac{\delta T}{\tau_\parallel}+\tilde{s}
\label{Eq:DampedBurgers}
\end{equation}
Here the magnetic drift $v_D$ is also introduced. Indeed, by balancing the magnetic drift and the parallel loss ($\sim1/\tau_\parallel$), we obtain the result of the heuristic drift model, namely that the width is set by $v_D\tau_\parallel\sim\epsilon_0\rho_\theta\sim\lambda_{HD}$. The question here is how avalanches can broaden the layer width beyond $\lambda_{HD}$. Note also that $\delta T$ is no longer a conserved order parameter due to the parallel damping. This limits the propagation of shocks in the damped Burgers model and implies that a sufficient drive is required for shock penetration, on which we elaborate below.

\begin{figure}[thb]
\centering\includegraphics[width=0.45\textwidth]{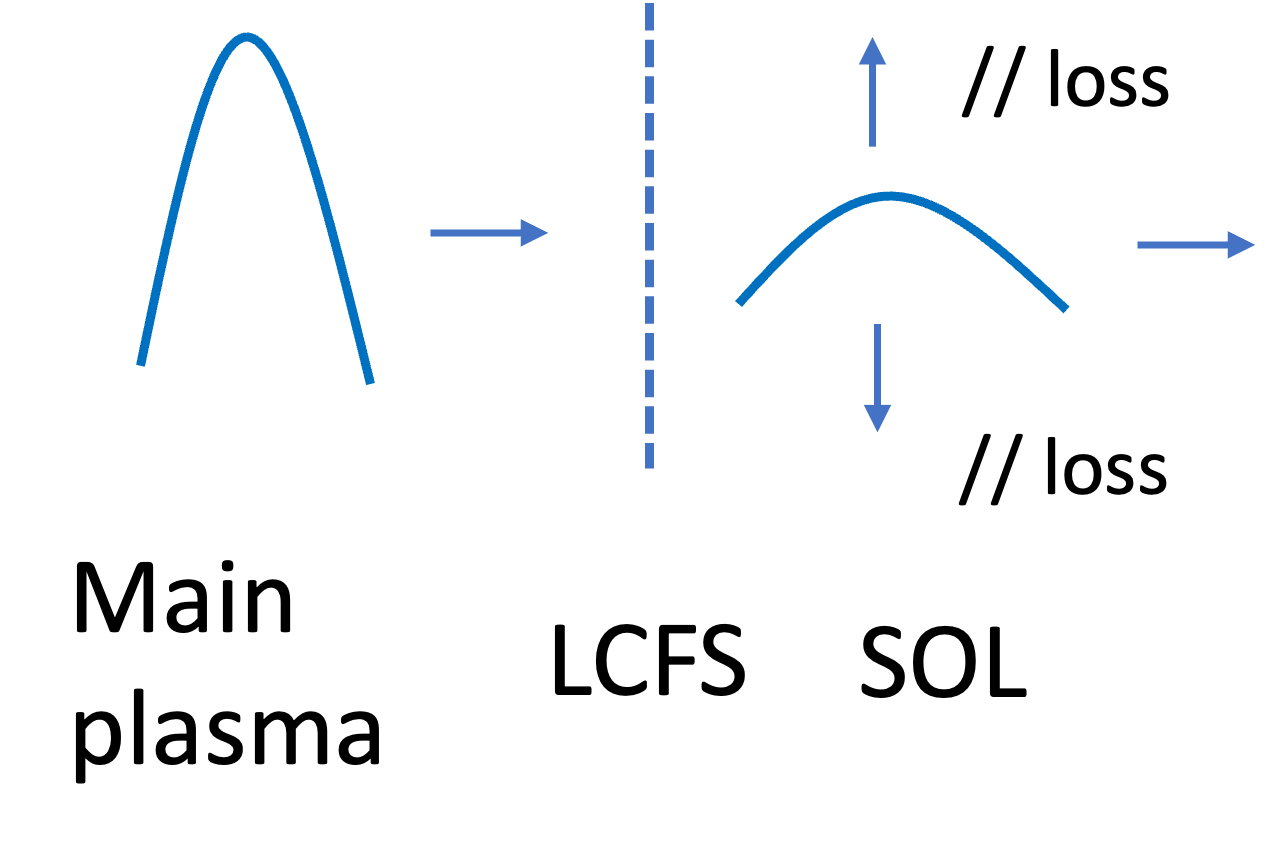}
\caption{Avalanches in the SOL suffer additional damping from the parallel loss.}
\label{Fig:SOLeffects}
\end{figure}

\begin{figure}[thb]
\centering\includegraphics[width=0.45\textwidth]{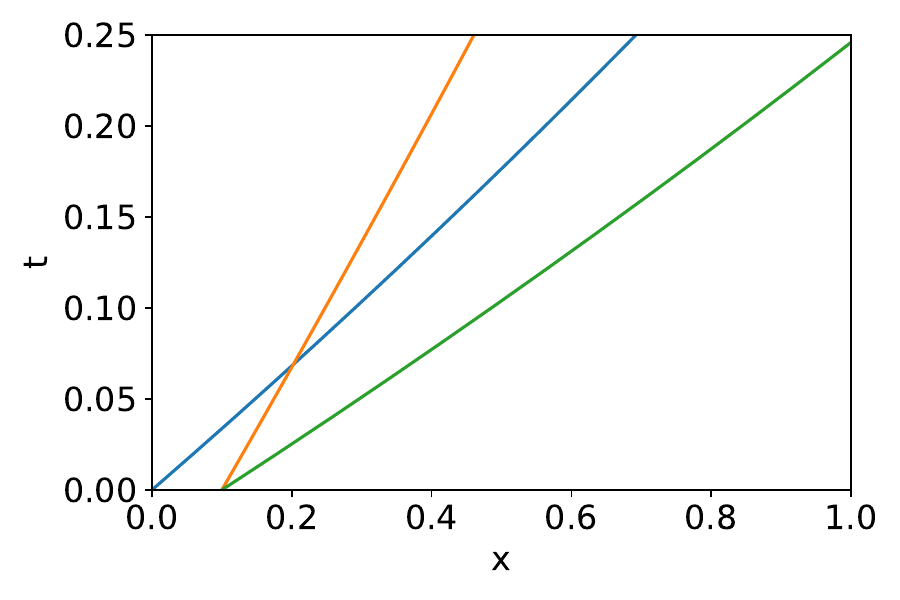}
\caption{Characteristics for damped Burgers equation. The yellow curve intersects the blue curve, at which point discontinuity forms and shock formation is expected. For the green curve, this is not the case. The condition for intersection, or shock formation, is set by the gradient of initial data (Eq.(\ref{Eq:ShockCondition})).}
\label{Fig:Characteristics}
\end{figure}

While avalanches in the SOL are attenuated by the parallel damping, shock formation is still possible in the damped Burgers model. To demonstrate this, here we discuss the condition for shock formation in the presence of the parallel damping\cite{Whitham}. This is done by using the method of characteristics, which rewrites Eq.(\ref{Eq:DampedBurgers}) as
\begin{subequations}
\begin{align}
&\frac{d}{dt}\delta T=-\frac{\delta T}{\tau_\parallel}
\label{Eq:ODE_Temp}\\
&\frac{dx}{dt}=v_{D}+\alpha \delta T
\label{Eq:ODE_Characteristics}
\end{align}
\end{subequations}
Thus avalanches decay exponentially along the characteristics as
\begin{equation}
\delta T(x,t)=\delta T(x=\xi,t=0)e^{-t/\tau_\parallel}
\end{equation}
$x(t=0)=\xi$ is an initial position and $\delta T(x=\xi,t=0)$ is an initial value of the temperature perturbation. The solution curve is also obtained as
\begin{equation}
x=\xi+v_Dt+\frac{1-e^{-t/\tau_\parallel}}{1/\tau_\parallel}f(\xi)
\end{equation}
where $f(\xi)=\alpha\delta T(x=\xi,t=0)$. Shock formation is indicated when characteristics intersect with each other. This occurs for
\begin{equation}
f'(\xi)<-1/\tau_\parallel\Leftrightarrow\alpha\frac{\partial \delta T(\xi,0)}{\partial \xi}<-1/\tau_\parallel
\label{Eq:ShockCondition}
\end{equation}
Thus the spatial gradient of initial perturbation needs to be steep enough to overcome the parallel damping. This effectively requires that the nonlinear term in the damped Burgers equation needs to overcome the parallel damping term. The case with and without shock formation is shown in Fig.\ref{Fig:Characteristics}. Here one characteristic (indicated by blue) starts from $(t,x)=(0,0)$ and two others (yellow and green) from $(t,x)=(0,0.1)$. For the latter, two different conditions are used; $\partial_\xi f(\xi)<-1/\tau_\parallel$ for yellow and $\partial_\xi f(\xi)>0$ for green. In these cases, the green curve has shallower inclination, and never intersects the blue curve. On the other hand, the yellow curve intersects the blue curve, indicative of discontinuity. These highlight the relevance of the condition (Eq.(\ref{Eq:ShockCondition})) for shock formation.

\section{Avalanche Penetration Study}
Having discussed a basic model for SOL avalanche dynamics and a condition for shock formation, here we turn to numerical analysis of the model. First we discuss our numerical setups, including relevant parameters after normalization. Then we present two different cases. One is a propagation study of a single pulse initiated in the region. The other addresses the impact of boundary forcing, which mimics the effect of avalanches incoming from the main plasma. Finally, we discuss the effectiveness of avalanche penetration in broadening SOL width.

\subsection{Numerical setup}
Equation (\ref{Eq:DampedBurgers}) is numerically solved here\cite{JamSim}. To do so, we first normalize the equation. Using $\tau_\parallel$ for the temporal scale, $\lambda_{HD}=\tau_\parallel v_D$ for the spatial scale, and $\delta T_0$ for a typical value of the temperature deviation (say $\alpha\delta T_0\sim v_*$), we have
\begin{subequations}
\begin{equation}
\frac{\partial}{\partial \hat{t}}\delta \hat T+\frac{\partial}{\partial \hat{x}}\delta \hat T+\hat{v}_{NL}\delta\hat{T}\frac{\partial}{\partial \hat{x}}\delta \hat T=\hat{\chi}_0\frac{\partial^2}{\partial \hat{x}^2}\delta \hat T-\delta\hat{T}
\label{Eq:DimlessBurgers}
\end{equation}
Here $\hat{(...)}$ denotes dimensionless variables. Noise term is dropped for simplicity, and forcing is provided from initial and boundary conditions, as specified below. Two remaining parameters are defined as
\begin{align}
&\hat{v}_{NL}=\frac{\alpha \delta T_0}{v_D}\\
\label{Eq:vnl}
&\hat{\chi}_0=\frac{\chi_0}{\tau_\parallel v_D^2}
\end{align}
\end{subequations}
Here the collisional conductivity is required for regulating shock steepening, so it will be kept as a small but finite value, i.e. $\hat{\chi}_0=0.01$ throughout this work. $\hat{v}_{NL}$ is the parameter that controls nonlinear steepening. Indeed, in this normalization, the shock condition (Eq.(\ref{Eq:ShockCondition})) is written as
\begin{equation}
\frac{\partial\delta\hat{T}(\hat{\xi},0)}{\partial \hat\xi}<-\frac{1}{\hat{v}_{NL}}
\label{Eq:ShockConditionDimLess}
\end{equation} 
The gradient of the temperature perturbation needs to overcome the critical value, which is set by $1/\hat{v}_{NL}$. Thus shock formation is likely for larger $\hat{v}_{NL}$, which gives a lower threshold. Although the derivative of the temperature gradient is more physical, $\hat{v}_{NL}$ is easier to scan in numerics. Then in the following we first present results for a parameter survey over $\hat{v}_{NL}$, and later elaborate its relation to a physical parameter, such as the gradient of temperature perturbation at the LCSF. Typical value of $\hat{v}_{NL}$ is $\hat{v}_{NL}\sim v_*/v_D\sim R/a\sim 3$, assuming that an avalanche starts penetrating from LCFS.  We perform parameter scans for $\hat{v}_{NL}$, ranging from linear case $\hat{v}_{NL}\ll1$ to a moderately nonlinear case $\hat{v}_{NL}=1\sim 5$. Hereafter, we drop $\hat{(...)}$ for notational simplicity.

\subsection{Propagation of an initial pulse}

\begin{figure}[thb]
\centering\includegraphics[width=0.45\textwidth]{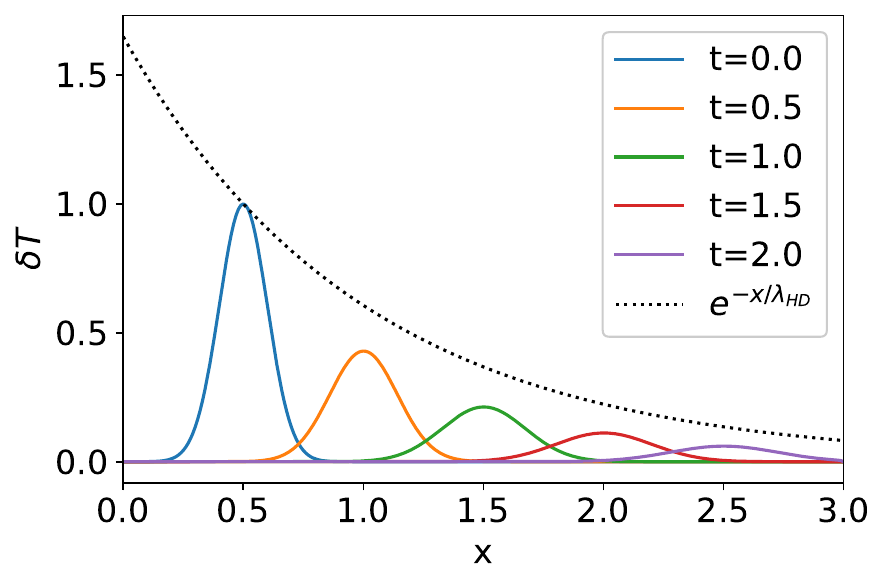}
\caption{Propagation of a localized pulse in damped Burgers model for weak drive, $v_{NL}=0.01$. This is effectively a linear limit in the model, and the pulse drifts, diffuses, and damps due to parallel loss.}
\label{Fig:DriftDiffusionLimit}
\end{figure}

We start our analysis by initiating a localized pulse in SOL region. Here we use a Gaussian pulse,
\begin{equation}
\delta T(t=0,x)=A_0\exp\left(-\frac{(x-x_0)^2}{2\Delta^2}\right)
\end{equation}
$A_0=1$ and $\Delta=0.1$ is used as a typical value. The location of pulse can be varied, and we start by initiating the pulse at $x_0=0.5$. As a reference case, we consider a linear limit, $\hat{v}_{NL}=0.01$. In this case, the shock condition is not satisfied, as the maximum value of the lefthand side of the Shock condition is $v_{NL}A_0/\Delta/\sqrt{e}\sim-0.06$. Equation (\ref{Eq:DampedBurgers}) is effectively linear, and straightforwardly solved as in Fig.\ref{Fig:DriftDiffusionLimit}. The initial pulse, located at $x_0=0$, drifts, diffuses, and then damps due to parallel diffusion. Penetration is shallow, and the decay scale tracks the heuristic drift model, $\lambda_{HD}$.

\begin{figure}[htb]
\centering\includegraphics[width=0.45\textwidth]{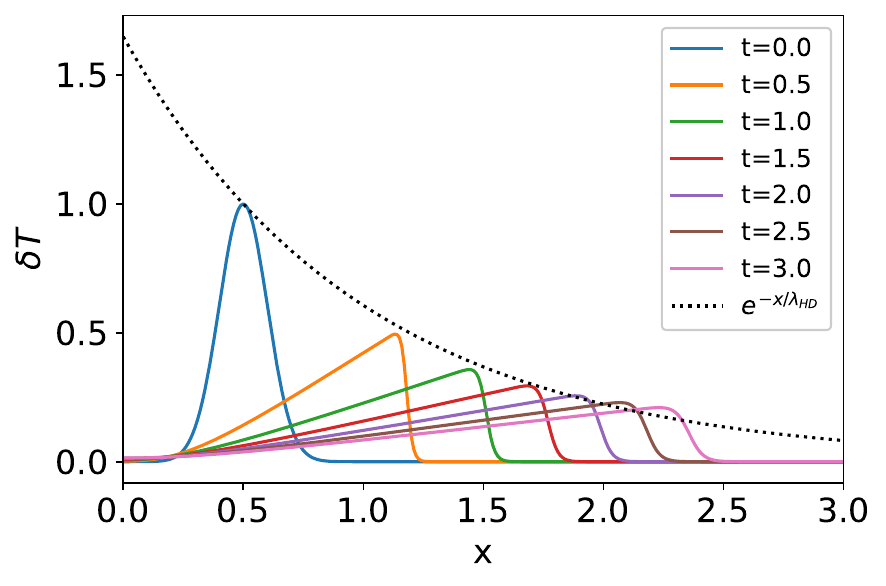}
\caption{Propagation of a localized pulse in the damped Burgers model for $v_{NL}=3$. The shock formation criterion is satisfied in this case. The pulse develops into a shock and penetrates deeper than the case for Fig.\ref{Fig:DriftDiffusionLimit}.}
\label{Fig:BlobShock}
\end{figure}

\begin{figure}[htb]
\centering\includegraphics[width=0.45\textwidth]{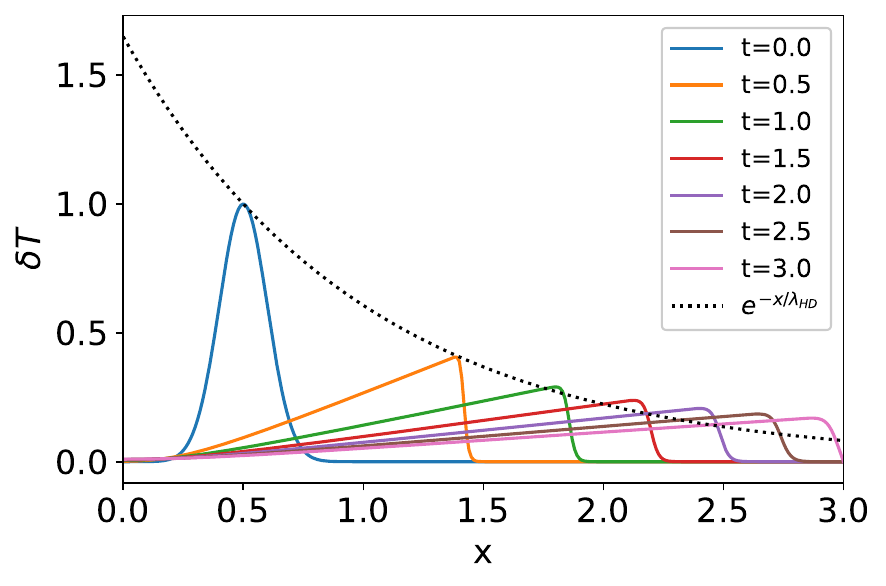}
\caption{Propagation of a localized pulse in the damped Burgers model for a stronger drive, $v_{NL}=5$. Shocks or avalanches penetrate deeper for larger $v_{NL}$.}
\label{Fig:BlobShockStronger}
\end{figure}

We then increase $v_{NL}$ to see the effect of shock formation on penetration. Fig.\ref{Fig:BlobShock} is the result for $v_{NL}=3$. Here, the shock condition is well-satisfied, since the LHS of the shock condition is $v_{NL}\partial \delta T/\partial \xi\sim-18$. As we can see in Fig.\ref{Fig:BlobShock}, the initial pulse steepens while it is propagating. Once a shock forms, it maintains its structure. We can further increase the nonlinear drive, and the result with $v_{NL}=5$ is shown in Fig.\ref{Fig:BlobShockStronger}. The penetration deepens as $v_{NL}$ increases. These demonstrate possible SOL widening by avalanche penetration.

\begin{figure}[htb]
\centering\includegraphics[width=0.45\textwidth]{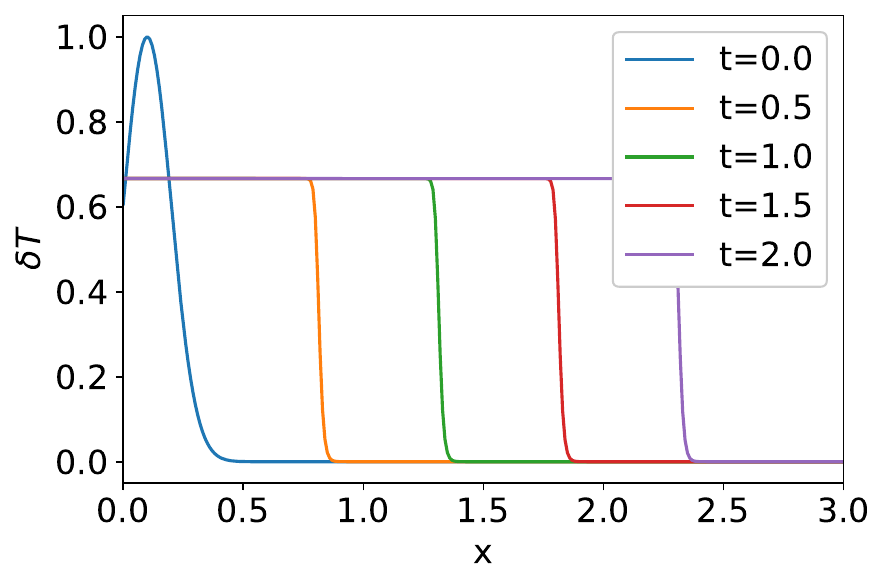}
\caption{A case with a propagating front in Burgers model}
\label{Fig:BurgersFront}
\end{figure}

\begin{figure*}[tbh]
\centering
\begin{subfigure}[b]{0.3\textwidth}
\includegraphics[width=\textwidth]{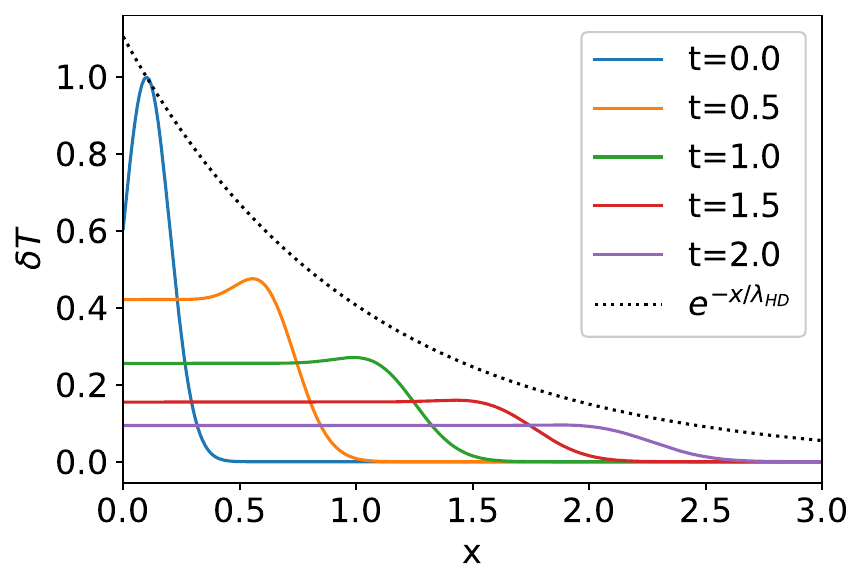}
\caption{$v_{NL}=0.01$}
\label{Fig:Front_HD}
\end{subfigure}
\centering
\begin{subfigure}[b]{0.3\textwidth}
\includegraphics[width=\textwidth]{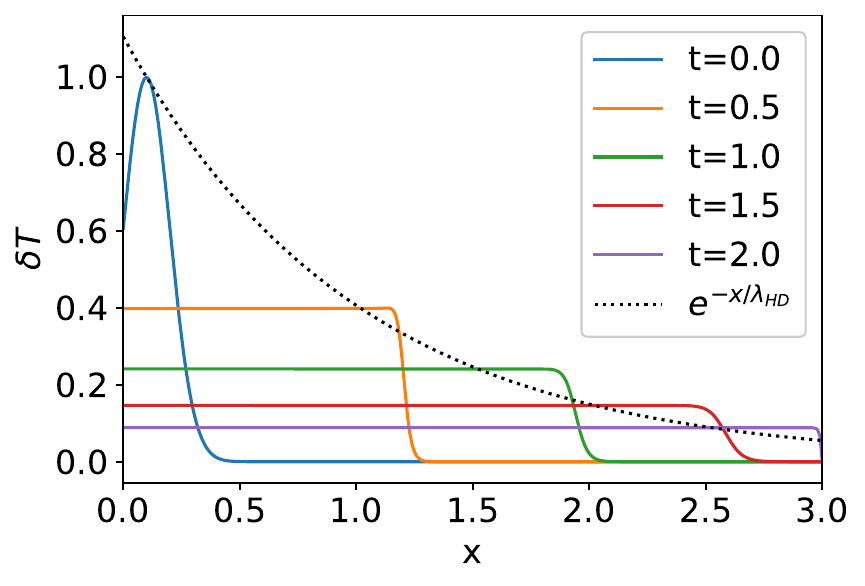}
\caption{$v_{NL}=3$}
\label{Fig:Front_NL}
\end{subfigure}
\centering
\begin{subfigure}[b]{0.3\textwidth}
\includegraphics[width=\textwidth]{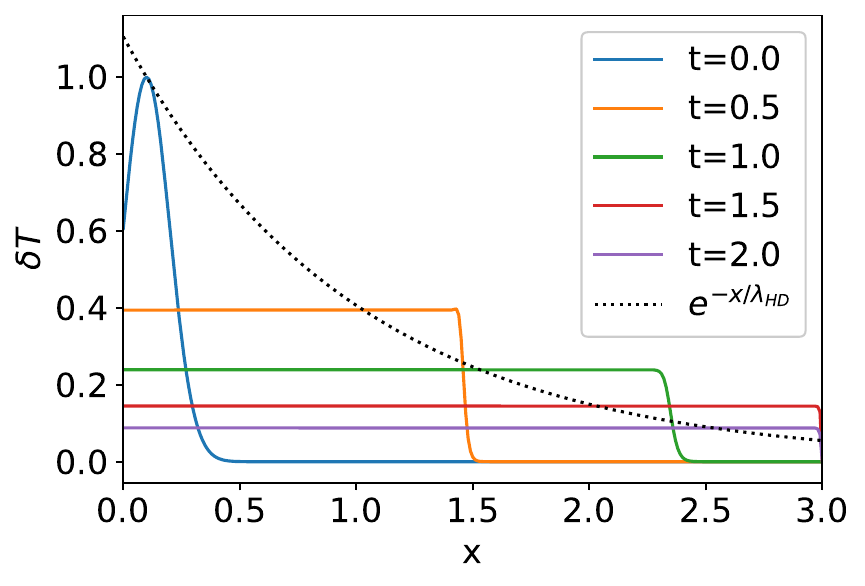}
\caption{$v_{NL}=5$}
\label{Fig:Front_NL_Stronger}
\end{subfigure}
\caption{Penetration of a front solution into SOL. Penetration is more pronounced than the cases shown for an isolated pulse. The penetration also becomes deeper for larger nonlinear drive, $v_{NL}$.}
\label{Fig:Front}
\end{figure*}

Shock propagation is more pronounced when we have a front, as shown in Fig.\ref{Fig:BurgersFront}. Here Fig.\ref{Fig:BurgersFront} is obtained via Burgers equation, to demonstrate its unique behavior. Unlike the case for an isolated pulse, here the pulse is initiated closer to the inner boundary. The boundary value remains finite, which acts as an effective forcing.  Comparing to the case of isolated pulse propagation in Burgers model in Fig.\ref{Fig:BlobVoid}, the front propagates faster and hence deeper. Thus we expect avalanches to penetrate deeper into SOL plasmas. This is indeed true, as shown in Fig.\ref{Fig:Front}. Fig.\ref{Fig:Front_HD} is the linear limit. In this case, while it initially propagates, a front damps very quickly due to parallel loss. With larger nonlinear drive, fronts maintain their form and penetrate into the SOL, as shown in Figs.\ref{Fig:Front_NL} and \ref{Fig:Front_NL_Stronger}. Indeed, they penetrate deeper than isolated pulses.

\subsection{Forced boundary}
\begin{figure}[tbh]
\centering\includegraphics[width=0.45\textwidth]{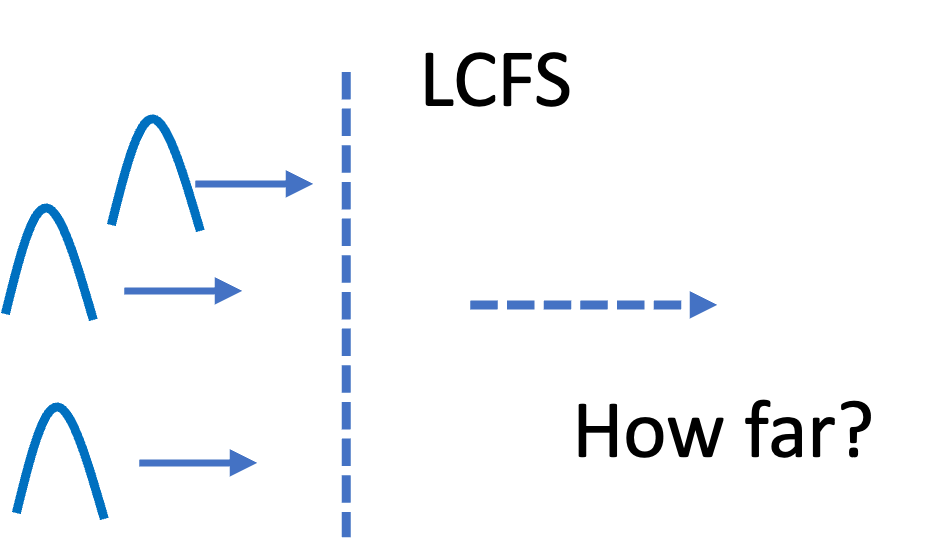}
\caption{Avalanches bombard the LCFS stochastically. This is mimicked by dynamic boundary forcing.}
\label{Fig:DynamicBoundary}
\end{figure}

We then consider the effect of boundary forcing. As shown in Fig.\ref{Fig:DynamicBoundary}, avalanches reach the LCFS in a stochastic manner. This leads to a stochastic forcing at the LCFS to initiate avalanche penetration. In this subsection, we discuss penetration of avalanches for noisy boundary forcing by noise.

\begin{figure*}[thb]
\centering
\begin{subfigure}[b]{0.45\textwidth}
\includegraphics[width=\textwidth]{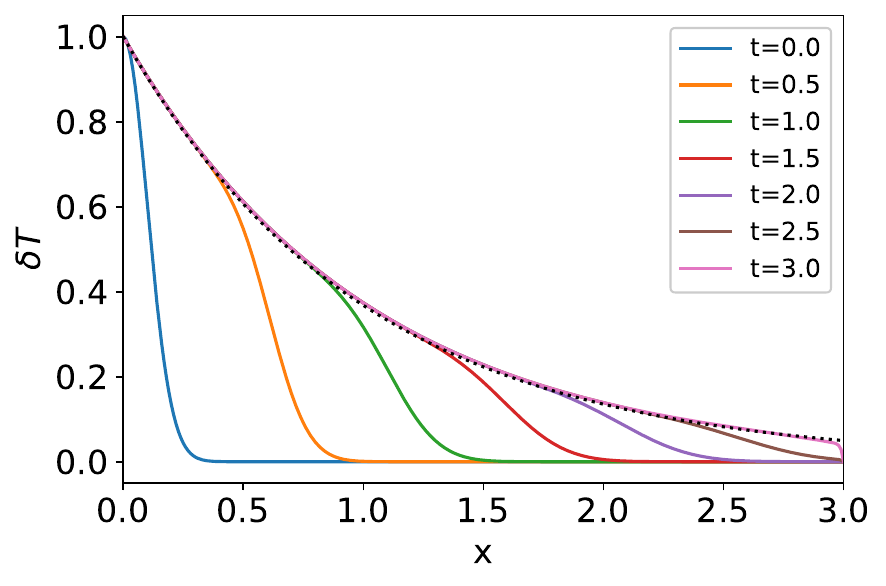}
\caption{$v_{NL}=0.01$}
\label{Fig:ConstantHD}
\end{subfigure}
\centering
\begin{subfigure}[b]{0.45\textwidth}
\includegraphics[width=\textwidth]{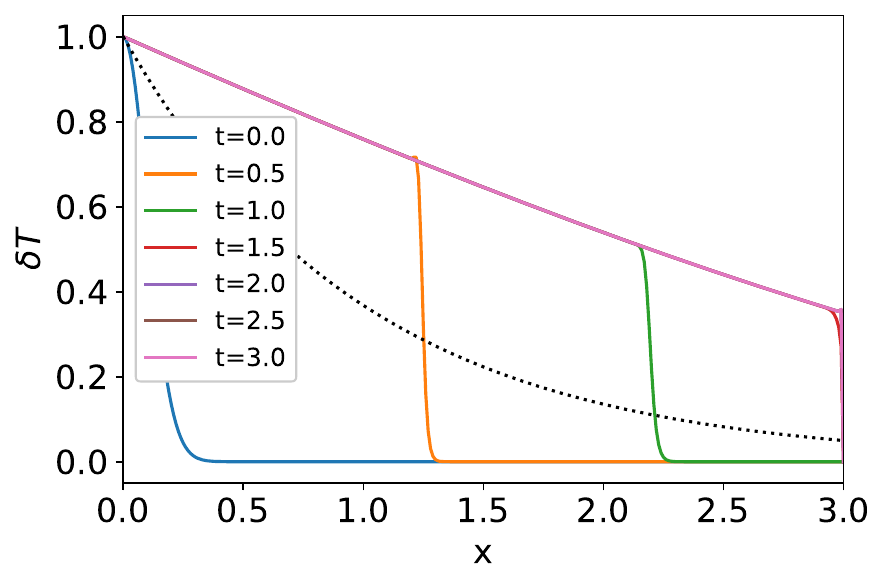}
\caption{$v_{NL}=3$}
\label{Fig:ConstantShock}
\end{subfigure}
\caption{A case with constant forcing at the inner boundary. Front propagation is evident for strong enough drive $v_{NL}$. The shock is very effective in broadening the profile.}
\label{Fig:ConstantB}
\end{figure*}

\begin{figure}[thb]
\centering\includegraphics[width=0.45\textwidth]{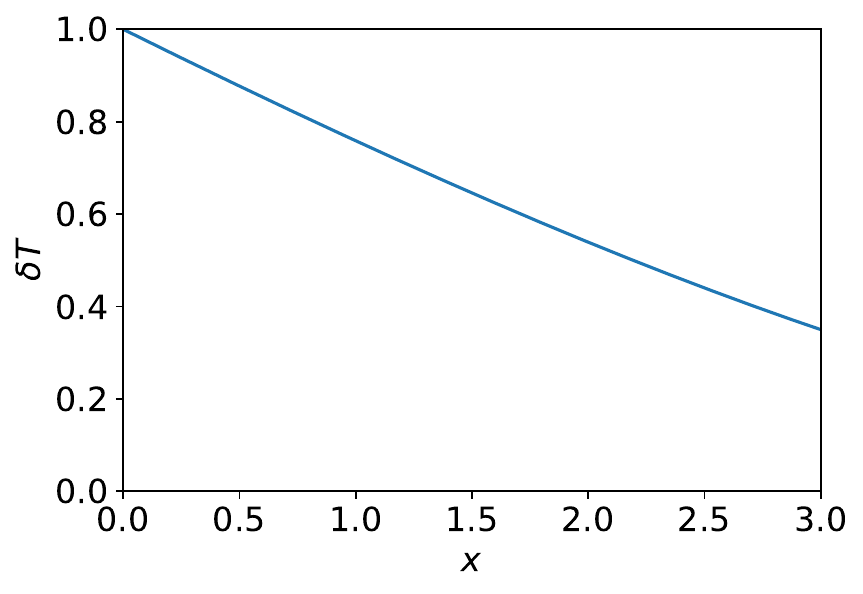}
\caption{A steady profile, set by Eq.(\ref{Eq:Profile}).}
\label{Fig:Profile}
\end{figure}

% Constant Forcing
The impact of the boundary forcing is shown explicitly by considering a case with constant injection. While our final goal is to understand the impact of dynamical injection of avalanches, we start with this simplified case. The result is shown in Fig. \ref{Fig:ConstantB}. Figure \ref{Fig:ConstantHD} is the linear case, where the profile is set by the magnetic drift and the parallel damping. The profile follows the heuristic-drift scaling, which is indicated by the black dash line. In contrast, the nonlinear case exhibits profile broadening, as shown in Fig.\ref{Fig:ConstantShock}. There the nonlinear term becomes relevant to the profile formation. Indeed, the steady state balance is given by
\begin{equation}
\partial_x\delta T+v_{NL}\delta T\partial_x\delta T=-\delta T
\end{equation}
This can be integrated to give
\begin{equation}
\log\frac{\delta T}{\delta T_0}+v_{NL}(\delta T-\delta T_0)=-x
\label{Eq:Profile}
\end{equation}
Here $\delta T_0=\delta T(x=0)$ and $\delta T_0=1$ for parameters for Fig.\ref{Fig:ConstantB}. The profile is plotted in Fig.\ref{Fig:Profile}, which shows a good agreement with the steady profile in Fig.\ref{Fig:ConstantShock}. We note that a slight deviation is seen close to the boundary, $x=3$, since the diffusion term is not included in Eq.(\ref{Eq:Profile}). We also note that the profile formation in the nonlinear case is similar to that of turbulent pipe flows. There, the linear limit of flow profile is set by the balance between the pressure drop along a channel and viscous dissipation. In the nonlinear (turbulent) case, the profile is broadened by turbulent momentum flux and a thin boundary layer forms to accommodate viscous dissipation. In that case, the nonlinear heat flux broadens temperature profile. Though simplified, this result demonstrates the relevance of boundary forcing and nonlinear drive for shock formation and profile broadening.

\begin{figure}[tbh]
\centering\includegraphics[width=0.45\textwidth]{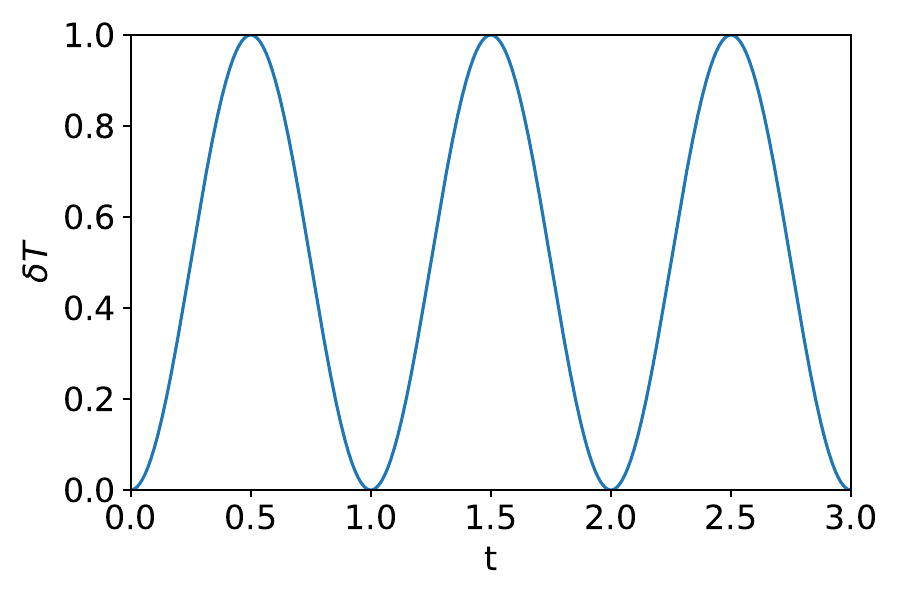}
\caption{The forcing at the inner boundary is dynamically changed. In this case, 3 avalanches are injected.}
\label{Fig:Boundary_N=3}
\end{figure}

\begin{figure*}[tbh]
\centering
\begin{subfigure}[b]{0.3\textwidth}
\includegraphics[width=\textwidth]{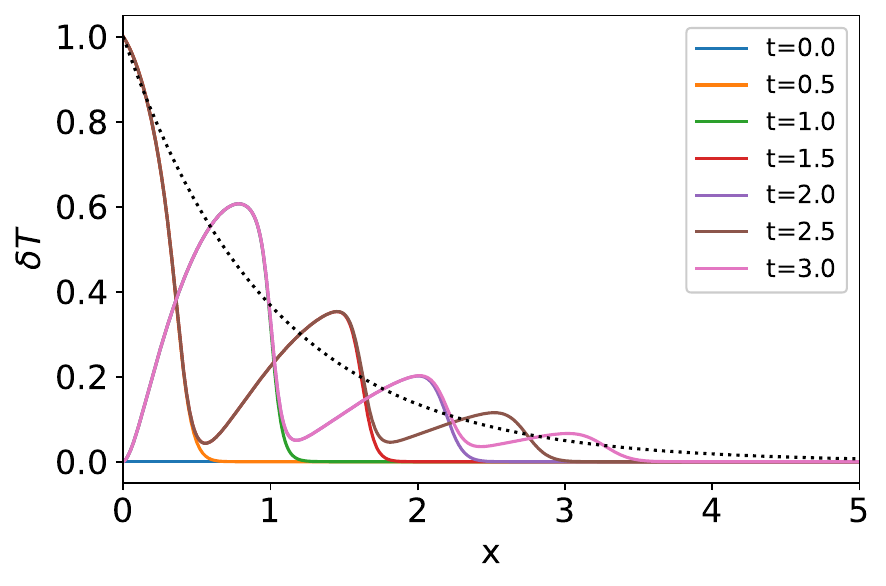}
\caption{$v_{NL}=1$}
\label{Fig:DBa}
\end{subfigure}
\centering
\begin{subfigure}[b]{0.3\textwidth}
\includegraphics[width=\textwidth]{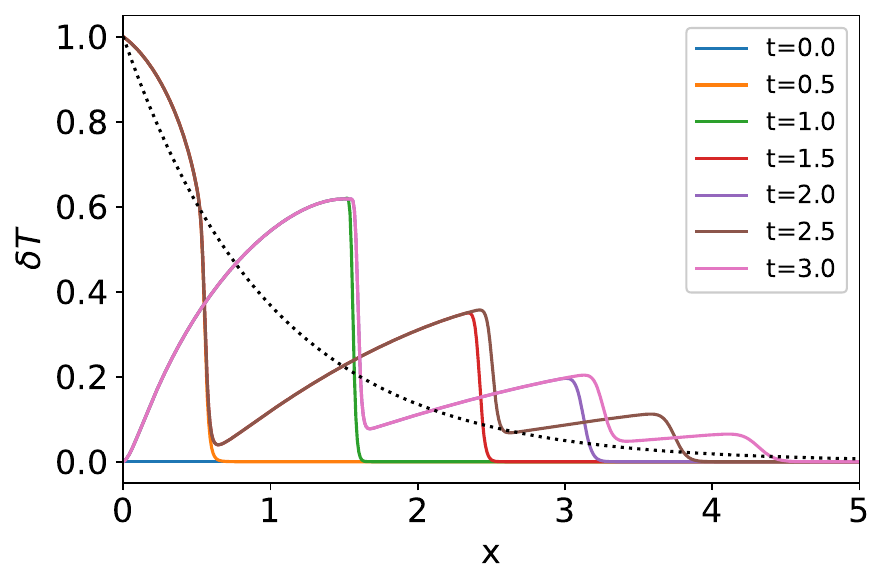}
\caption{$v_{NL}=3$}
\label{Fig:DBb}
\end{subfigure}
\centering
\begin{subfigure}[b]{0.3\textwidth}
\includegraphics[width=\textwidth]{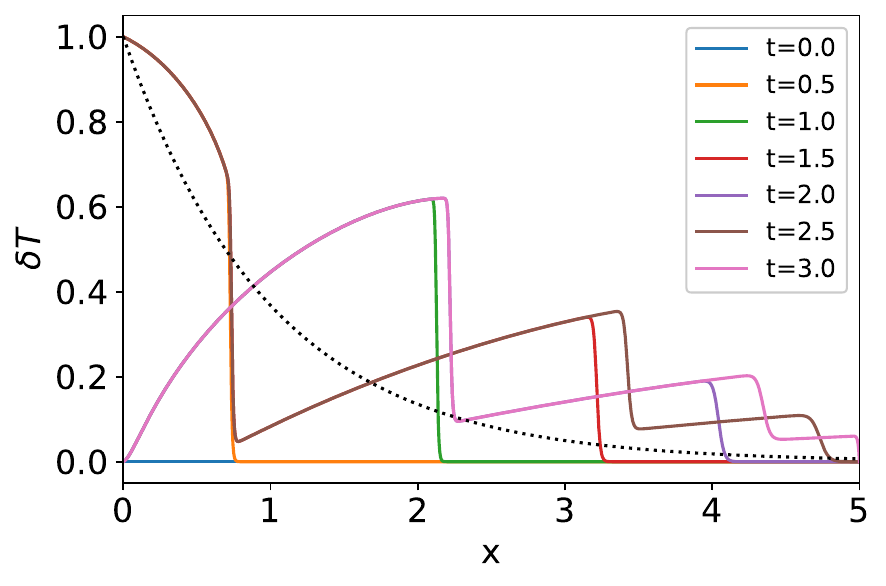}
\caption{$v_{NL}=5$}
\label{Fig:DBc}
\end{subfigure}
\caption{Penetration of avalanches by dynamic forcing. As confirmed in $t=3$, three avalanches are penetrating in this case, corresponding to the boundary forcing (Fig.\ref{Fig:Boundary_N=3}).}
\label{Fig:DB}
\end{figure*}

% Dynamic forcing
Since the LCFS is stochastically bombarded by avalanches (Fig.\ref{Fig:DynamicBoundary}), it is of interest to analyze the case with dynamic boundary forcing. We mimic the effect of the dynamic injection by varying  the forcing at the inner boundary in time, as shown in Fig.\ref{Fig:Boundary_N=3}. This case corresponds to the injection of three avalanches. Avalanches penetrate into the SOL, as shown in Fig.\ref{Fig:DB}. Here three cases are shown for moderately nonlinear forcing. At later stage of the penetration, say $t=3$ which is indicated by pink, we can identify 3 pulses or avalanches. Avalanches penetrate deeper, as $v_{NL}$ increases (Figs.\ref{Fig:DBb} and \ref{Fig:DBc}). Dynamically injected avalanches can broaden the SOL profile.

\subsection{Penetration Depth}
\begin{figure}[tbh]
\centering\includegraphics[width=0.45\textwidth]{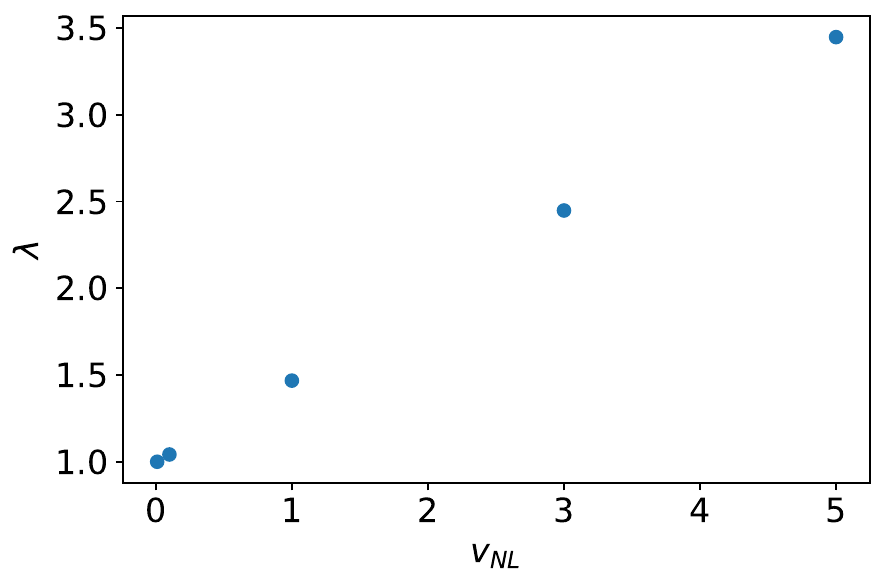}
\caption{Dependence of the penetration depth, a proxy to the SOL width, on the drive parameter ($v_{NL}$) for shocks.}
\label{Fig:DecayLength}
\end{figure}

As demonstrated so far, avalanches can penetrate into the SOL plasma and the penetration becomes deeper for larger NL drive ($v_{NL}$). This effect of avalanche penetration can be made more quantitative by analyzing the dependence of the penetration depth as a function of the nonlinear drive. The penetration depth is evaluated by exponential fitting. The decay length is calculated, and the results for the dynamic injection case (3 avalanches) is shown in Fig.\ref{Fig:DecayLength}. Here the decay length is normalized to the reference value, obtained for the linear limit $v_{NL}=0$. As we can see, there is a positive correlation, and as the nonlinear drive increases, the decay length, a proxy to the SOL width, increases. Recalling that $v_{NL}$ is directly tied to the condition for shock formation, this result shows that \textit{the shock formation and avalanche penetration are favorable for broadening the SOL.}

% Data analysis
It is of great interest here to derive a typical criterion that indicates penetration of avalanches. To do so, we recall that  the shock formation condition (in physical units) requires
\begin{equation}
\left|\frac{\partial \delta T}{\partial r}\right|>\left|\frac{\partial \delta T}{\partial r}\right|_{crit}=\frac{1}{\alpha\tau_\parallel}.
\end{equation}
In dimensionless units, this translates as
\begin{equation}
\left|\frac{\partial\delta\hat{T}(\hat{\xi},0)}{\partial \hat\xi}\right|>\left|\frac{\partial\delta\hat{T}(\hat{\xi},0)}{\partial \hat\xi}\right|_{crit}=\frac{1}{\hat{v}_{NL}}
\end{equation}
where $\hat{v}_{NL}=\alpha\delta T_0/v_{D}$. Although numerical studies above indicate penetration of avalanches for large $\hat{v}_{NL}$, which corresponds to a lower threshold, this condition can be also satisfied for large $|\partial_r\delta T_{LCFS}|$. Since the latter is more physical, and since $\delta T/T$ is measurable in experiments, here we estimate a typical value of $\delta T/T$ for avalanche penetration. Recalling that $\alpha$ is related to a typical speed of avalanches $V(\delta T)\sim\alpha \delta T$ and assuming that $V(\delta T)$ is of the order of the dynamic diamagnetic velocity, set by the gradient of the temperature perturbation, we have
\begin{equation}
V(\delta T)\sim\alpha\delta T\sim\tilde{v}_{dia}\sim\frac{\rho_sc_s}{T_0}\frac{\partial\delta T}{\partial r}.
\end{equation}
Then the coefficient $\alpha$ is estimated to be
\begin{equation}
\alpha \sim\frac{\rho_sc_s}{T_0}\frac{\partial}{\partial r}\sim\frac{\rho_sc_s}{T_0}k_r.
\end{equation}
Here note that the radial derivative is replaced by a typical wave number of avalanches, $k_r$. Thus, the lefthand side of the shock formation criterion scales as
\begin{equation}
\alpha\frac{\partial \delta T}{\partial r}\sim\rho_sc_s k_r^2\frac{\delta T}{T_0}\sim\omega_{ci}\rho_s^2 k_r^2\frac{\delta T}{T_0}.
\end{equation}
This needs to overcome the damping from the parallel loss, so we have
\begin{subequations}
\begin{equation}
\omega_{ci}\rho_s^2 k_r^2\frac{\delta T}{T_0}>\frac{1}{\tau_\parallel},
\end{equation}
or in terms of amplitude,
\begin{equation}
\frac{\delta T}{T_0}>\frac{1}{\omega_{ci}\tau_\parallel\rho_s^2 k_r^2}\sim\frac{\rho_s}{qR}\frac{1}{k_r^2\rho_s^2}.
\end{equation}
Here it is understood that the criterion is evaluated at separatrix. Taking $\Delta_r\sim k_r^{-1}$ as a typical scale of avalanche at the separatrix, we finally have
\begin{equation}
\left.\frac{\delta T}{T_0}\right|_{sep}>\frac{\Delta_r^2}{\rho_sRq}.
\label{Eq:CritAmp}
\end{equation}
\end{subequations}
The scale of avalanches is typically in the range of the several correlation length. In the region of interest, $\Delta_r$ is at most in the order of the pedestal width, $\Delta_r\sim\Delta_{ped}\sim\rho_\theta$. Then we typically have $\Delta_r/(qR)\sim\rho_\theta/(qR)\sim\rho_s/(\epsilon_0R)\sim\rho_*\sim10^{-2} - 10^{-3}$ and $\Delta_r/\rho_s\sim\rho_\theta/\rho_s\sim q/\epsilon\sim 10$. Then we find the required amplitude for penetration as $\delta T/T_0|_{sep}\sim O(0.01) - O(0.1)$. We would expect this condition can be satisfied at a turbulent edge, allowing the penetration of avalanches into SOL (Fig.\ref{Fig:WidthTemp}). In particular, for an extreme avalanche event, this condition can be well satisfied and will allow deep penetration. Of course this is somewhat speculative, and more careful analysis is desirable, using stochastic forcing effects with a distribution of avalanching events. This will be pursued in a future publication.

\section{Summary and discussion}
In this work, we introduced a simplified model for avalanche propagation in the SOL and presented numerical analyses of the model. The principal results are as follows.
\begin{enumerate}
\item Avalanche penetration into the SOL is modeled via a damped Burgers equation (Eq.\ref{Eq:DampedBurgers}). The parallel loss is included as a Krook operator. In this model, the temperature perturbation is no longer a conserved order parameter. Thus avalanches attenuate. A critical drive is required for penetration.
\item The condition for avalanches to penetrate into the SOL is derived in terms of a condition for shock formation in the damped Burgers equation. The condition (Eq.\ref{Eq:ShockCondition}) requires strong enough drive (gradient of temperature perturbation) to overcome the parallel damping. When this condition is satisfied, a shock forms, to allow avalanches both to penetrate the SOL, and to broaden the SOL width, as shown schematicaly in Fig.\ref{Fig:WidthTemp}.
\item Numerical analysis demonstrates the relevance of shock formation for avalanche penetration. Here we presented results for a scan of the parameter $v_{NL}$, which is a coefficient of the nonlinear term in the dimensionless damped Burgers equation (Eq.\ref{Eq:DimlessBurgers}). A larger $v_{NL}$ corresponds to a lower threshold for shock formation. In particular,
	\begin{enumerate}
	\item An isolated, localized pulse is initiated and its propagation is calculated. When the drive is weak, avalanches decay quickly, and the penetration depth is aligned with the heuristic drift model (e.g. Fig.\ref{Fig:DriftDiffusionLimit}). On the other hand, when nonlinear drive is sufficient, shocks form and penetrate the SOL (Fig.\ref{Fig:BlobShock}). The penetration depth increases as the nonlinear drive parameter ($v_{NL}$) increases (Fig. \ref{Fig:BlobShockStronger}).
	\item A case with a propagating front is investigated. In contrast to the case with an isolated pulse, a front penetrates deeper (Fig.\ref{Fig:Front}). SOL profile broadening is also demonstrated with constant heating and front propagation (Fig.\ref{Fig:ConstantShock}).
	\item Avalanches stochastically bombard the LCFS and this effect is modeled by a dynamic boundary condition. Multiple avalanches are found to penetrate (Fig.\ref{Fig:DB}).
	\item The decay length is calculated by fitting the avalanche amplitude by an exponential damping. A positive correlation is found between the penetration depth and the driving parameter $v_{NL}$ (Fig.\ref{Fig:DecayLength}).
	\end{enumerate}
\item An estimate is given for critical perturbation amplitude for penetration (Eq.(\ref{Eq:CritAmp})). The shock formation condition requires $|\partial_r\delta T|_{rms, sep}>|\partial_r\delta T|_{crit}=1/(\alpha\tau_\parallel)\propto 1/v_{NL}$. To connect to measurement,  this condition is rewritten in terms of the avalanche amplitude, $(\delta T/T_0)|_{sep}>\Delta_r^2/(\rho_sRq)$, where $\Delta_r$ is a typical radial extent of an avalanche. For typical parameters of interest, we found $(\delta T/T_0)|_{sep}\sim O(0.1)-O(0.01)$ is required for penetration. When the condition is satisfied, penetration of avalanches into SOL is expected (Fig.\ref{Fig:WidthTemp}).
\end{enumerate}

% some victory
In short, heat avalanches penetrate the SOL when they can shock. A sufficient drive is necessary for shock formation against the parallel loss. The gradient of temperature perturbation emerges as a prime candidate for this. When the gradient is below a threshold, shocks do not form, and avalanches quickly attenuate in the SOL. Once the gradient exceeds a critical value, avalanches penetrate the SOL. We note that this is consistent with experiments on Heliotron J, where the radial correlation of the temperature fluctuation across the LCFS becomes finite for high heating power.

% scaling analysis
Although avalanche penetration into the SOL has been demonstrated in this work, there is room for further improvements and extension. These include, but are not limited to, the radial electric field, two dimensionality, electromagnetic fluctuation, etc. In particular, a relevant issue is to demonstrate avalanche penetration using data from more realistic simulations and experiments, such as BOUT++ data or probe data from experiments. Along with the penetration condition discussed in this work, we can look for footprints of avalanches in the heat flux dynamics in SOL region. The correlation between the SOL width and separatrix temperature perturbation gradients may be explored. These will be discussed in future works.

\acknowledgments
We ackowledge stimulating discussions with F. Kin, Zeyu Li, and participants of the JIFT workshop held in San Diego, 2025. This work is partly supported by; the Grants-in-Aid for Scientific Research of JSPS of Japan (JP21H01066, JP23K20838, JP17H06089); the joint research project in RIAM; the U.S. Department of Energy, Office of Science, Office of Fusion Energy Sciences under Award No. DE-FG02-04ER54738; the Sci DAC ABOUND Project, scw1832; the EPSRC under Grant No. EP/R014604/1.

\bibliographystyle{apsrev4-1}
\bibliography{Penetration}

\end{document}